\documentclass[]{JHEP3} 

\usepackage{epsfig,multicol,multirow,bbm,bm}

\newcommand\fverb{\setbox\fverbbox=\hbox\bgroup\verb}
\newcommand\fverbdo{\egroup\medskip\noindent%
            \fbox{\unhbox\fverbbox}\ }
\newcommand\fverbit{\egroup\item[\fbox{\unhbox\fverbbox}]}
\newbox\fverbbox

\title{One-dimensional holographic superconductor from AdS$_3$/CFT$_2$ correspondence}

\author{Jie Ren\\
Department of Physics, Princeton University\\
Princeton, NJ 08544, U.S.A.\\
E-mail: \email{jieren@princeton.edu}}

\preprint{PUPT-2347}    

\abstract{We obtain a holographical description of a superconductor by using the $d=2$ case of the AdS$_{d+1}$/CFT$_d$ correspondence. The gravity system is a (2+1)-dimensional AdS black hole coupled to a Maxwell field and charged scalar. The dual (1+1)-dimensional superconductor will be strongly correlated. The characteristic exponents for vector perturbations at the boundary degenerate, which implies that $d=2$ is a critical dimension and the Green's function needs to be regularized. In the normal phase, the current-current correlation function and the conductivity can be analytically solved at zero chemical potential. The dc conductivity can be analytically solved at finite chemical potential. When we add a scalar hair to the black hole, a charged condensate happens at low temperatures. We compare our results with higher-dimensional cases.}

\keywords{AdS-CFT Correspondence, Black Holes, Field Theories in Lower Dimensions}

\begin{document}


\section{Introduction}
AdS/CFT (anti-de Sitter/conformal field theory) correspondence \cite{mal97} enables us to study some strongly coupled quantum field theories by means of general relativity, and this approach provides new universality classes of condensed matter systems. For example, the (2+1) and (3+1)-dimensional holographic superconductors have been achieved by AdS$_4$/CFT$_3$ and AdS$_5$/CFT$_4$ correspondences \cite{gub08,har08a,har08b,hor08}. By applying the AdS/CFT correspondence, the correlation functions in the CFT at the boundary can be extracted from a classical gravity system in the bulk \cite{gub98,wit98,son02}. Various aspects of the AdS$_3$/CFT$_2$ correspondence have also been studied. The Ba\~{n}ados-Teitelboim-Zanelli (BTZ) black hole \cite{ban92} is often taken to be the gravity part. It has been shown that the quasinormal modes in this spacetime coincide with the poles of the correlation function in the dual CFT, which gives quantitative evidence for AdS/CFT \cite{bir02}. In Ref.~\cite{mai10}, the charged BTZ black hole is taken to be the gravity part, as a generalization to break the scale invariance. Some other gravity systems from superstring theory are also used
\cite{dav09,hun10}. However, there is no superconducting phase transition in previous works.

We explore the (1+1)-dimensional holographic superconductor using the AdS$_3$/CFT$_2$ correspondence and show its distinctive features in both normal and superconducting phases. We calculate the conductivity, which is obtained from the current-current correlation function. At first sight, the conductivity is not well-defined in AdS$_3$. If we add $e^{-i\omega t}A_x(z)$ as a perturbation, the asymptotic behavior near the boundary is $A_x=\alpha+\beta z^{d-2}$ in AdS$_{d+1}$ ($d>2$). It can be chosen that $\alpha$ is the source, and $\beta$ is the expectation value. Thus, the Green's function is defined as $G\sim\beta/\alpha$. In AdS$_3$, the asymptotic behavior is $A_x={\cal A}\ln z+{\cal B}$, because the characteristic exponents degenerate. The Green's function $G\sim{\cal B}/{\cal A}$ is used in previous works \cite{mai10,hun10}. When $z\to 0$, the leading term is ${\cal A}\ln z$, which implies that ${\cal A}$ is the source at the boundary. And ${\cal B}$ is the expectation value after we add a counterterm to regularize the divergence. Consequently, the Green's function is defined as $G=-{\cal B}/{\cal A}$. Another point of view is that $d=2$ is a critical dimension and the Green's function can be regularized by considering $d=2+\epsilon$ dimensions.

The real-time correlation functions of vector and tensor perturbations can be obtained by using gauge invariant variables \cite{kov05}. The full current-current correlation function from AdS$_3$ is analytically solvable at finite temperature and zero chemical potential. Thus, we obtain the frequency and momentum dependence of the conductivity $\sigma(\omega,k)$, which describes the linear response to the both temporally and spatially oscillating electric field. The conductivity has poles at $\omega=k$, and there is no diffusive mode in AdS$_3$. We take the charged AdS$_3$ black hole as the gravity dual to calculate the frequency dependence of the conductivity at finite chemical potential. The conductivity can be analytically solved in the dc limit, and thus we obtain the temperature dependence of the dc conductivity. The charged AdS$_3$ black hole corresponds to the normal phase of the holographic superconductor. At low temperatures, an instability will happen and cause the black hole to develop scalar hair \cite{gub08,har08b}.

To achieve the superconducting phase, we add a scalar field, and calculate the conductivity numerically. We mainly study the case that the mass of the scalar field equals the Breitenlohner-Freedman (BF) bound \cite{bre82}. We find that the one-dimensional holographic superconductor can be realized by the gravity dual where the charged black hole develops scalar hair. Below a critical temperature $T_c$, the real part of the conductivity Re[$\sigma(\omega)$] has a delta function at $\omega=0$ and an apparent gap $\omega_g$, similar to higher-dimensional cases \cite{har08a,har08b,hor08}. We also check that the Ferrell-Glover-Tinkham (FGT) sum rule is obeyed in both normal and superconducting phases. Near $T_c$, the scalar operator behaves similarly to that in BCS theory. But the superfluid density, which is related to the imaginary part of the conductivity, appears to be divergent at $T=T_c$. We argue that this divergence can be regularized by the dimension $d=2+\epsilon$.

This paper is organized as follows. In section 2, we write down the gravitational dual of the one-dimensional holographic superconductor, and discuss the definition of the Green's function and conductivity. In section 3, we solve the current-current correlation function and obtain the frequency and momentum dependence of the conductivity at zero chemical potential. In section 4, we study the normal phase of the superconductor at finite chemical potential. In section 5, we add the scalar hair and explore the superconducting phase numerically. In section 6, we discuss the superconductor away from the probe limit, and the conductivity at zero temperature. In section 7, we conclude with a summary and some open questions.

\section{Setup for AdS$_3$/CFT$_2$}
The minimal Lagrangian of the gravitational dual describes a charged black hole and scalar field in asymptotically AdS spacetime. The action is
\begin{equation}
S=\int
d^{d+1}x\sqrt{-g}\left(R+\frac{d(d-1)}{L^2}-\frac{1}{4}F_{\mu\nu}F^{\mu\nu}-|\nabla\Psi-iqA\Psi|^2-V(\Psi)\right),\label{eq:action}
\end{equation}
where $V(\Psi)=m^2|\Psi|^2$. The instability to form scalar hair is analyzed in Sec. 6. Before we reach the superconducting phase, we will start from the pure AdS$_3$, and then add temperature, chemical potential, and finally the scalar hair step by step in the following sections. The metric ansatz of Poincar\'{e} coordinates is
\begin{equation}
ds^2=\frac{L^2}{z^2}\left(-f(z)dt^2+dx^2+\frac{dz^2}{f(z)}\right).\label{eq:ansatz}
\end{equation}
The conformal boundary is at $z=0$, and the horizon is at $z=z_h$, where $f(z_h)=0$. We set $L=1$ and $z_h=1$,
which are allowed by two scaling symmetries \cite{har08b}. The solution to the equations of motion corresponds
to a thermal equilibrium state at the boundary.

The asymptotic behavior of a field $\Theta$ near the boundary is
\begin{equation}
\Theta(z)={\cal A}z^{\Delta_-}(1+\cdots)+{\cal B}z^{\Delta_+}(1+\cdots),
\end{equation}
where $\Delta_\pm$ ($\Delta_-<\Delta_+$) are characteristic exponents of the perturbation equation, and the
dots are terms that vanish as $z\to 0$. For a scalar field, $\Delta_\pm$ are solutions to
$\Delta(\Delta-d)=m^2L^2$, while for a vector field, $\Delta_\pm$ are solutions to $\Delta(\Delta-d+2)=m^2L^2$.
Thus, for a massless vector field, $\Delta_-=0$ and $\Delta_+=d-2$. Usually, ${\cal A}$ corresponds to the
source of an operator ${\cal O}$ at the boundary, and ${\cal B}$ corresponds to its expectation value
$\langle{\cal O}\rangle$. The retarded Green's function at the boundary corresponds to $\langle{\cal
OO}\rangle_R\sim\cal B/\cal A$, with incoming-wave boundary condition at the horizon. If the two scaling
dimensions degenerate, there will be a logarithm and we need to examine more carefully how the Green's function
is defined. Note that if $\Delta_+-\Delta_-=2n$ ($n=1,2,\cdots$), there is also a logarithm, but this logarithm
does not affect the identification of the source, only introducing an ambiguity in the Green's function. In the
following, we will use the current-current correlation function to show how to define the Green's function.

To obtain the conductivity, we need to perturb the system by adding $e^{-i\omega t}A_x(z)$. The Green's function is defined from the variation of the action as $G^{xx}=J^x/A_x$, where $J^x$ is the conserved current measuring the linear response with respect to an external electric field. Ohm's law gives $J^x=\sigma^{xx}E_x=i\omega\sigma^{xx}A_x$. Therefore, we obtain the conductivity from the Green's function by $\sigma^{xx}=G^{xx}/i\omega$. We often simply denote $G^{xx}$ and $\sigma^{xx}$ as $G$ and $\sigma$, respectively. Consider the case without the scalar field. To take into account the backreaction of the Maxwell field on the metric, we also need to add $e^{-i\omega t}g_{tx}(z)$. Then we obtain two linearized equations for $A_x$ and $g_{tx}$ in the background Eq.~(\ref{eq:ansatz}). After eliminating $g_{tx}$, we obtain a single equation for $A_x$ as
\begin{equation}
A_x''+\left(\frac{f'}{f}-\frac{d-3}{z}\right)A_x'+\left(\frac{\omega^2}{f^2}-\frac{A_t'^2z^2}{f}\right)A_x=0,\label{eq:pertAx}
\end{equation}
where the prime denotes derivative with respect to $z$. When $d=2$, both the characteristic exponents of this
equation at the boundary $z=0$ are zero. Therefore, the asymptotic behavior of $A_x$ is
\begin{equation}
A_x={\cal A}\ln(\Lambda z)+{\cal B}+\cdots,
\end{equation}
where $\Lambda$ is a renormalization scale included in the logarithm, and the dots are terms that vanish as
$z\to 0$. The leading term is ${\cal A}\ln z$, which implies that ${\cal A}$ is the source. We will show that
the following prescription gives the Green's function:
\begin{equation}
G=-\frac{\cal B}{\cal A}.\label{eq:green}
\end{equation}
There will be no minus sign if we use another coordinate $r=1/z$ and write $A_x$ as $A_x={\cal
A}\ln(\Lambda r)+{\cal B}+\cdots$. The Green's function has an ambiguity due to the logarithmic term, i.e., it can
be shifted by a constant $\ln\Lambda$.

The boundary conditions for the vector field are studied in detail in Ref.~\cite{mar06}. We start with the
action for the vector field with a surface term:
\begin{equation}
S=-\frac{1}{4}\int d^3x\sqrt{-g}F_{\mu\nu}F^{\mu\nu}+\int d^2x\sqrt{-\gamma}n_\mu A_\nu
F^{\mu\nu}.
\end{equation}
If $A_x={\cal A}\ln z+{\cal B}$, the variation of the action is
\begin{equation}
\delta S=\int d^2x\sqrt{-\gamma}n^\mu A^\nu\delta F_{\mu\nu}=\int
d^2x\sqrt{-\gamma}g^{zz}g^{xx}n_zA_x\partial_z(\delta A_x)=\int d^2x({\cal A}\ln z+{\cal B})\delta{\cal A},
\end{equation}
where $\gamma_{\mu\nu}$ is the induced metric on the boundary, and $n^\mu$ is the inward normal vector of the
boundary. The $\ln z$ divergence can be regularized by adding the counterterm \cite{hun10}
\begin{equation}
S_{\rm ct}=\frac{1}{2\ln\Lambda}\int d^2x\sqrt{-\gamma}A_\mu A_\nu\gamma^{\mu\nu},
\end{equation}
and evaluating the total action at $z=1/\Lambda$. Thus, we can see that ${\cal A}$ corresponds to the source and ${\cal B}$ corresponds to the expectation value.

It turns out that the Green's function can also be obtain by taking the $d\to 2$ limit of $d=2+\epsilon$. If $d>2$, the Green's function is defined as follows. The asymptotic behavior near the boundary is $A_x=\alpha+\beta z^{d-2}$. Starting with the action without the surface term, we obtain
\begin{equation}
\delta S=-\int d^dx\sqrt{-\gamma}n^\mu F_{\mu\nu}\delta A^\nu=-\int
d^dx\sqrt{-\gamma}g^{zz}g^{xx}n_z(\partial_zA_x)\delta A_x=-\int d^dx(d-2)\beta\delta\alpha.
\end{equation}
When $d>2$, this is a well-defined variation at the boundary:
\begin{equation}
\frac{\delta S}{\delta(\epsilon\alpha)}=-\langle\beta\rangle,
\end{equation}
where $\epsilon=d-2$, and $\alpha$ corresponds to the source. Therefore, the Green's function is defined as
\begin{equation}
G=\frac{\beta}{\epsilon\alpha}.\label{eq:greenab}
\end{equation}
Instead of $G=\epsilon\beta/\alpha$, this comes from an alternative normalization and gives nontrivial results when $\epsilon\to 0$ \cite{kle99}.

In the gravity part, we consider the pure AdS$_{d+1}$ for simplicity, since the Green's function can be exactly
solved for arbitrary $d$ in this case. The perturbation equation for $A_x$ is
\begin{equation}
A_x''-\frac{d-3}{z}A_x'+\omega^2A_x=0.\label{eq:Axbdy}
\end{equation}
Considering the boundary conditions, the solution is
\begin{equation}
A_x\sim z^{d/2-1}H^{(1)}_{d/2-1}(\omega z),
\end{equation}
where $H_\nu^{(1)}(z)$ is the Hankel function of the first kind. By expanding $A_x$ near $z=0$ as
$A_x=\alpha+\beta z^{d-2}+\cdots$, we obtain the Green's function
\begin{equation}
G(\omega)=\frac{2^{2-d}\pi\omega^{d-2}(i-\cot(d\pi/2))}{(d-2)\Gamma(d/2)\Gamma(d/2-1)}.\label{eq:greend}
\end{equation}
When $d$ is even, it contains a pole, which should be subtracted. If we expand Eq.~(\ref{eq:greend}) around
$d=2$, we obtain
\begin{equation}
G=-\frac{1}{\epsilon}+\left(-\gamma+\frac{\pi i}{2}-\ln\frac{\omega}{2}\right)+{\cal
O}(\epsilon),\label{eq:green2}
\end{equation}
where $\epsilon=d-2$. By subtracting the pole, taking into account the ambiguity $\Lambda$, and using
$\sigma=G/i\omega$, we obtain the conductivity
\begin{equation}
\sigma(\omega)=\frac{1}{\omega}\left(\frac{\pi}{2}+i\left(\gamma+\ln\frac{\omega}{2\Lambda}\right)\right),\label{eq:zeroT}
\end{equation}
This result is obtained by regarding the AdS$_3$ as a limit of AdS$_{d+1}$, where $d\to 2$ and with a
renormalization.

On the other hand, we can apply our prescription Eq.~(\ref{eq:green}) to obtain the Green's function. The
asymptotic behavior of $A_x$ for AdS$_3$ is
\begin{equation}
A_x\sim H_0^{(1)}(\omega z)=1+\frac{2i}{\pi}\left(\gamma+\ln\frac{\omega}{2}\right)+\frac{2i}{\pi}\ln z+{\cal
O}(z^2).
\end{equation}
The conductivity obtained by Eq.~(\ref{eq:green}) is the same as Eq.~(\ref{eq:green2}) after subtracting the
pole.

By the way, we give the general result of the conductivity calculated from Eq.~(\ref{eq:greend}):
\begin{equation}
\sigma(\omega)=\left\{\begin{array}{ll}\displaystyle\frac{\omega^{d-3}}{[(d-2)!!]^2}, & \qquad d=3,5,\cdots\\
\displaystyle\frac{\omega^{d-3}}{[(d-2)!!]^2}\left(\frac{\pi}{2}+i\ln\frac{\omega}{\Lambda_d}\right), & \qquad
d=2,4,\cdots\end{array}\right.\label{eq:sigads}
\end{equation}
where $\Lambda_d$ is a renormalization scale that has absorbed other constants. This result is essentially determined by the conformal symmetry. In particular, $\sigma=1$ for $d=3$, which is also true for AdS$_4$ at finite temperature \cite{her07a}. When $d$ is even, there is an ambiguity in the imaginary part.

\section{Current-current correlation function and conductivity}
Conductivity is obtained from the current-current correlation function, which depends on both frequency and
momentum in general. The frequency and momentum dependent conductivity describes the linear response to a both
temporally and spatially oscillating electric field. The retarded Green's function is
\begin{equation}
\tilde{G}_{\mu\nu}(x-y)=i\theta(x^0-y^0)\langle[J_\mu(x),J_\nu(y)]\rangle,
\end{equation}
where $J_\mu$ is the conserved current. The Fourier transform is denoted by $G_{\mu\nu}(p)$, where
$p^2=-\omega^2+{\bm k}^2$. At zero temperature, all components of $G_{\mu\nu}$ are determined by a scalar function $\Pi(p^2)$ as $G_{\mu\nu}=(\eta_{\mu\nu}-p_\mu p_\nu/p^2)\Pi(p^2)$. At finite temperature, the Lorentz invariance is broken, and $G_{\mu\nu}$ can be split into transverse and longitudinal parts:
\begin{equation}
G_{\mu\nu}(p)=P_{\mu\nu}^T\Pi^T(\omega,{\bm
k})+P_{\mu\nu}^L\Pi^L(\omega,{\bm k}).
\end{equation}
In AdS$_3$/CFT$_2$, there is only one spatial dimension at the boundary, and thus no transverse channel, so $\Pi=\Pi^L(\omega,k)$. The relation among components of the correlation function is \cite{kov05}
\begin{equation}
G_{tt}=\frac{k^2}{\omega^2-k^2}\Pi^L,\qquad
G_{tx}=G_{xt}=-\frac{\omega k}{\omega^2-k^2}\Pi^L,\qquad
G_{xx}=\frac{\omega^2}{\omega^2-k^2}\Pi^L.
\end{equation}

To obtain the full correlation function by AdS/CFT, we have to turn on $e^{-i\omega t+ikx}A_t(z)$ and
$e^{-i\omega t+ikx}A_x(z)$ as perturbations. The linearized Maxwell equations are
\begin{eqnarray}
\omega A_t'+kfA_x' &=& 0,\label{eq:AxAt}\\
A_t''+\frac{1}{z}A_t'-\frac{k^2}{f}A_t-\frac{\omega k}{f}A_x &=& 0,\\
A_x''+\left(\frac{f'}{f}+\frac{1}{z}\right)A_x'+\frac{\omega^2}{f^2}A_x+\frac{\omega k}{f^2}A_t &=& 0,
\end{eqnarray}
only two of which are independent. From the above equations, we can obtain a single equation for the gauge
invariant variable $E_x=i(kA_t+\omega A_x)$ \cite{kov05} as
\begin{equation}
E_x''+\left(\frac{\omega^2f'}{(\omega^2-k^2f)f}+\frac{1}{z}\right)E_x'+\frac{\omega^2-k^2f}{f^2}E_x=0,\label{eq:vector}
\end{equation}
where $f(z)=1-z^2$. The correlation function can be obtained by solving $E_x$ with incoming-wave boundary condition at the
horizon. If $d\geq 3$, $E_x$ can be expanded as $E_x={\cal A}+{\cal B}z^{d-2}+\cdots$ near the boundary $z=0$, and
then the Lorentzian AdS/CFT prescription gives $\Pi^L\sim{\cal B}/{\cal A}$ \cite{son02,kov05}. If $d=2$, the
situation is similar to what we discussed in the previous section. To obtain the Green's function $G^{xx}$, we
have to rewrite the action in terms of the gauge invariant variable $E_x$, and examine its variation. By
expanding $E_x$ as $E_x={\cal A}\ln(\Lambda z)+{\cal B}+\cdots$, the longitudinal self-energy $\Pi^L$ is obtained by
\begin{equation}
\Pi^L=-\frac{\cal B}{\cal A}.\label{eq:PiL}
\end{equation}
This can also be justified by taking the $d\to 2$ limit of pure AdS$_{d+1}$, in which $f(z)=1$ and the $1/z$
term in Eq.~(\ref{eq:vector}) should be replaced by $-(d-3)/z$. The solution of $E_x$ is
\begin{equation}
E_x\sim z^{d/2-1}H^{(1)}_{d/2-1}(\sqrt{\omega^2-k^2}z).
\end{equation}
After solving $E_x$, we obtain all components of the correlation function from $\Pi^L$.

It turns out that Eq.~(\ref{eq:vector}) can be solved in terms of hypergeometric functions. If $d\geq 3$, the
full correlation function is not analytically solvable, although it can be calculated in the hydrodynamical limit, i.e., when $\omega$ and $k$ are small \cite{pol02}. However, the correlation function is analytically solvable in AdS$_3$. The solution of
Eq.~(\ref{eq:vector}) can be written as
\begin{equation}
E_x(z)=C_1(1-z^2)^{-i\omega/2}y(z)+C_2(1-z^2)^{i\omega/2}y^*(z),
\end{equation}
where
\begin{eqnarray}
y(z) &=& (1-z^2)z^2\Big(\frac{\omega^2-k^2}{4}+i\omega-1\Big)
\,_2F_1\left(2-\frac{i(\omega+k)}{2},2-\frac{i(\omega-k)}{2},2-i\omega,1-z^2\right)\nonumber\\
&&
+\Big(\frac{2-i\omega}{2}z^2-1\Big)(i\omega-1)\,_2F_1\left(1-\frac{i(\omega+k)}{2},1-\frac{i(\omega-k)}{2},1-i\omega,1-z^2\right),
\end{eqnarray}
where $_2F_1(a,b,c;z)$ is the hypergeometric function. The incoming-wave boundary condition at the horizon
requires $C_2=0$. Then we expand $E_x$ as $E_x={\cal A}\ln(\Lambda z)+{\cal B}+\cdots$, and we obtain
$\Pi^L=-{\cal B}/{\cal A}$. To recover the explicit temperature dependence, we replace $z$ and $\omega$ with
$z/z_h$ and $\omega/(2\pi T)$, respectively, where $T=1/(2\pi z_h)$. The result is
\begin{equation}
\Pi^L(\omega,k)=-\frac{2\pi iT\omega}{\omega^2-k^2}-\frac{1}{2}\left[2\gamma+\psi\left(-\frac{i(\omega+k)}{4\pi
T}\right)+\psi\left(-\frac{i(\omega-k)}{4\pi T}\right)\right]+\ln\frac{\Lambda}{2\pi T},\label{eq:PiLok}
\end{equation}
where $\psi$ is the digamma function defined by $\psi(z)=\Gamma'(z)/\Gamma(z)$, and $\gamma$ is Euler's
constant. Here $\gamma$ is not important since it can be eliminated by a rescaling of $\Lambda$. The poles of
the Green's function are quasinormal nodes at
\begin{equation}
\omega\pm k=-4\pi iTn,\qquad n=0,1,2,\cdots
\end{equation}
which was obtained in Ref.~\cite{bir02}. There is no diffusive mode with a pole at $\omega=-iDk^2$, which exists in higher-dimensional cases. We can obtain the ``hydrodynamical limit" directly by expanding $E_x$ in terms of small $\omega$ and $k$:
\begin{equation}
E_x\sim-(\omega^2-k^2)\ln z+i\omega+\cdots,
\end{equation}
from which we obtain
\begin{equation}
\Pi^L=\frac{i\omega}{\omega^2-k^2}.
\end{equation}

\begin{figure}
\centering
\begin{minipage}[c]{0.5\textwidth}
\centering
\includegraphics[width=0.95\textwidth]{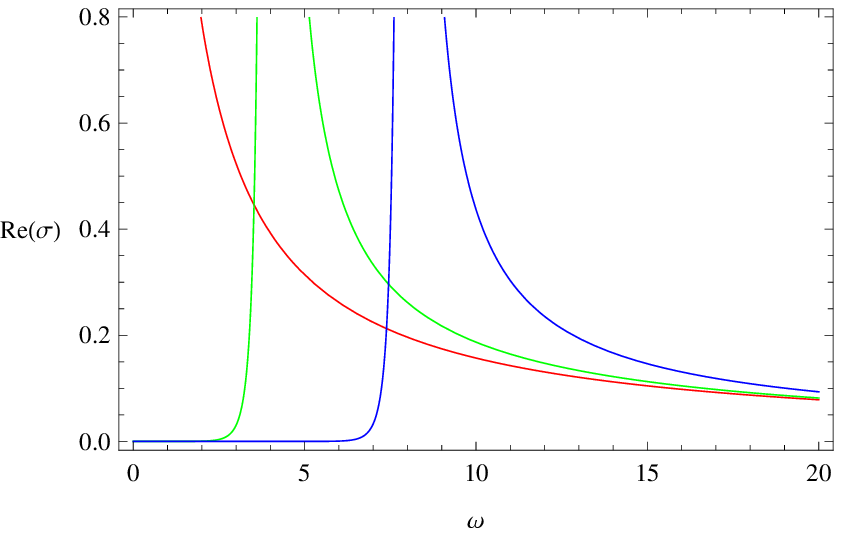}
\end{minipage}%
\begin{minipage}[c]{0.5\textwidth}
\centering
\includegraphics[width=0.95\textwidth]{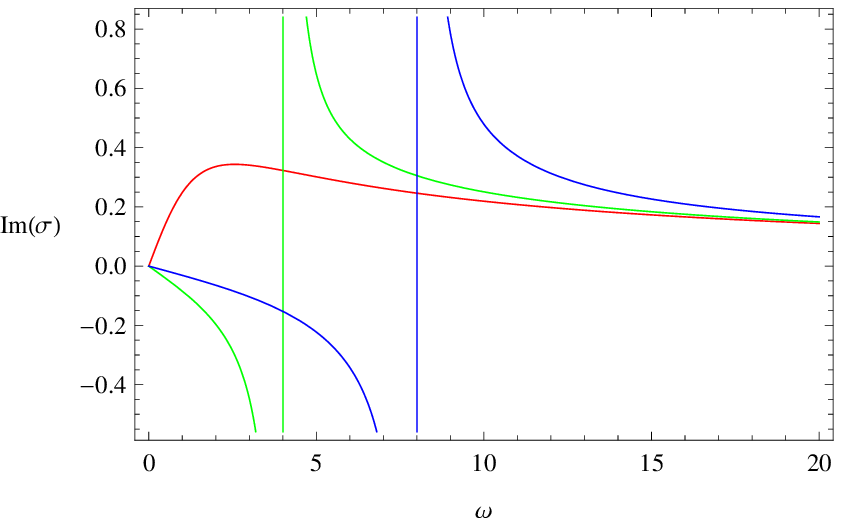}
\end{minipage}
\caption{\label{fig:sigmaT} Real and imaginary parts of the conductivity calculated from AdS$_3$ at finite
temperature and zero chemical potential. The horizon is at $z_h=1/(2\pi T)=1$. And $\Lambda=1$. The red, green, and blue
lines are the frequency dependence of the conductivity at momentum $k=0$, 4, 8, respectively.}
\end{figure}

The relation between the Green's function and conductivity is as follows. Ohm's law gives
\begin{equation}
J^x=\sigma E_x=i\sigma(kA_t+\omega A_x).
\end{equation}
On the other hand,
\begin{equation}
J^x=G^{xx}A_x+G^{xt}A_t=\frac{\omega\Pi^L}{\omega^2-k^2}(\omega A_x+kA_t),
\end{equation}
where $G^{xt}=-G_{xt}$. By comparing the above two equations, we obtain the conductivity
\begin{equation}
\sigma=\frac{\omega\Pi^L}{i(\omega^2-k^2)}=\frac{G^{xx}}{i\omega}.
\end{equation}
The real and imaginary parts of the conductivity $\sigma(\omega,k)$ are plotted in Fig.~\ref{fig:sigmaT}. We
compare the conductivity at zero and finite momentum. From the plots, we can see that the large $\omega$
behaviors are the same. When $\omega<k$, the real part Re($\sigma$) becomes very small.

We take $k=0$ and focus on the frequency dependence of the conductivity in the rest of this paper. We can
obtain the result from Eq.~(\ref{eq:PiLok}) by setting $k=0$. But we will obtain the conductivity from $A_x$ in
the following. By adding a perturbation $e^{-i\omega t}A_x(z)$, the linearized equation for $A_x$ is
\begin{equation}
A_x''+\left(\frac{f'}{f}+\frac{1}{z}\right)A_x'+\frac{\omega^2}{f^2}A_x=0,
\end{equation}
where $f=1-z^2$. The solution with incoming-wave boundary condition at the horizon $z=1$ is
\begin{equation}
A_x\sim(1-z^2)^{-i\omega/2}\,_2F_1\left(1-\frac{i\omega}{2},-\frac{i\omega}{2},1-i\omega,1-z^2\right).
\end{equation}
Then we expand this solution near the boundary $z=0$ as $A_x=\mathcal{A}+\mathcal{B}\ln(\Lambda z)+\cdots$. Consequently, the conductivity is given by
\begin{equation}
\sigma(\omega)=\frac{G(\omega)}{i\omega}=\frac{i}{\omega}\left[\gamma+\frac{2\pi iT}{\omega}+\psi\left(-\frac{i\omega}{4\pi
T}\right)\right]-\frac{i}{\omega}\ln\frac{\Lambda}{2\pi T}.\label{eq:sigT}
\end{equation}
Two asymptotic behaviors of the conductivity are as follows. For $\omega<<T$,
\begin{equation}
\sigma(\omega)=\frac{2\pi T}{\omega^2}-\frac{i}{\omega}\ln\frac{\Lambda}{2\pi
T}+\frac{\pi}{24T}-\frac{i\psi''(1)\omega}{32\pi^2T^2}+{\cal O}(\omega^2),\label{eq:smallo}
\end{equation}
where $\psi''(z)$ is the second derivative of the digamma function. And for $\omega>>T$,
\begin{equation}
\sigma(\omega)=\frac{1}{\omega}\left(\frac{\pi}{2}+i\left(\gamma+\ln\frac{\omega}{2\Lambda}\right)\right)+{\cal
O}(\omega^{-3}).\label{eq:largeo}
\end{equation}
The leading order is independent of $T$ and consistent with the zero temperature case given by
Eq.~(\ref{eq:zeroT}). We can see that the $\omega\to 0$ and $T\to 0$ limits do not commute.

A similar gravity system (built from intersecting D3-branes) with a (1+1)-dimensional dual was studied in Ref.~\cite{hun10}. The asymptotic behavior of the conductivity obtained in Ref.~\cite{hun10} is similar to the leading order terms in Eqs.~(\ref{eq:smallo}) and (\ref{eq:largeo}). The authors demonstrate their system resembles a Luttinger liquid \cite{voi95} in certain aspects.

\section{Normal phase}
\begin{figure}
\centering
\begin{minipage}[c]{0.5\textwidth}
\centering
\includegraphics[width=0.95\textwidth]{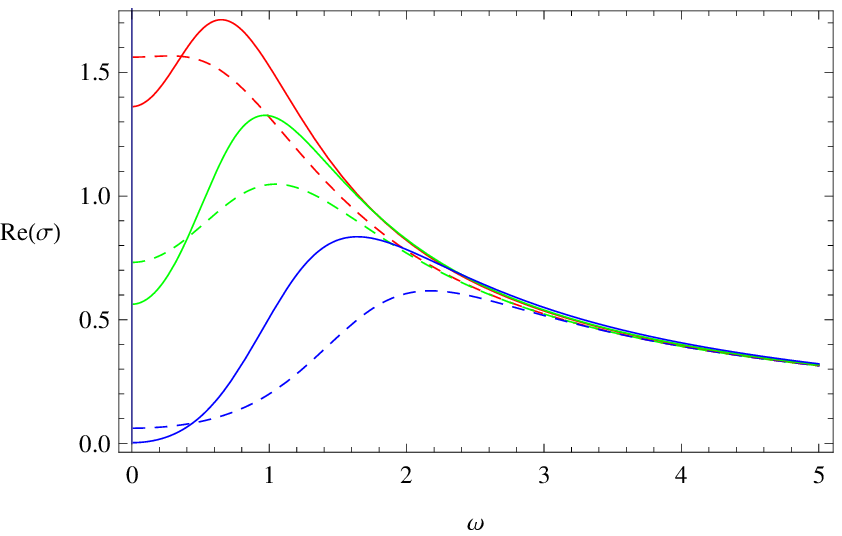}
\end{minipage}%
\begin{minipage}[c]{0.5\textwidth}
\centering
\includegraphics[width=0.95\textwidth]{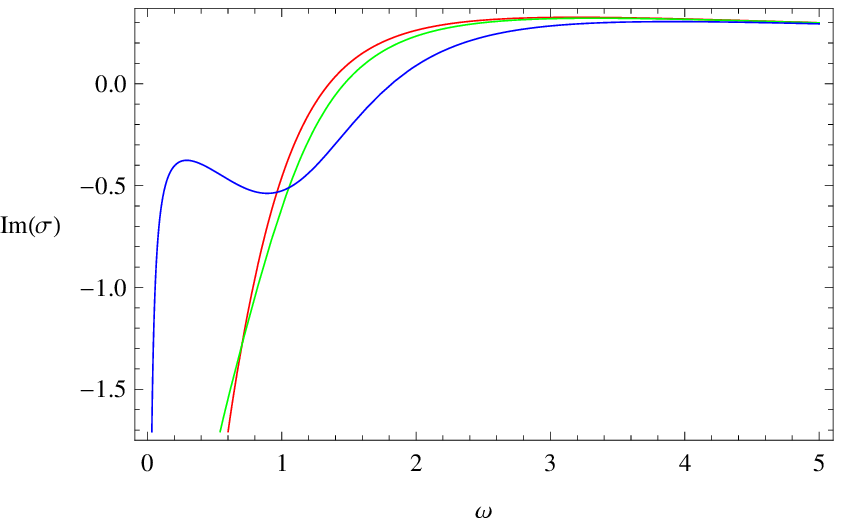}
\end{minipage}
\caption{\label{fig:sq} Real and imaginary parts of the conductivity calculated from AdS$_3$ at finite temperature and finite chemical potential. The horizon is at $z_h=1$. The red, green, and blue solid lines are the frequency dependence of the conductivity at $2\pi T=0.7$, 0.6, 0.1, i.e., $\mu^2=1.2$, 1.6, 3.6, respectively. The dashed lines are the real part of the approximate analytic result at $\mu^2=1.2$, 1.6, 3.6.}
\end{figure}

The normal phase of the superconductor corresponds to a charged AdS$_3$ black hole without scalar hair. The
solution of the Einstein-Maxwell equations with the metric ansatz Eq.~(\ref{eq:ansatz}) and $A=\phi(z)dt$ is
given by
\begin{eqnarray}
f(z) &=& 1-\frac{z^2}{z_h^2}+\frac{\mu^2z^2}{2}\ln\frac{z}{z_h},\\
A(z) &=& \mu\ln\frac{z}{z_h}\,dt,
\end{eqnarray}
where $\mu$ is the chemical potential, and $z_h$ is the position of the horizon. This is the non-rotating BTZ black hole. The Hawking temperature is
\begin{equation}
T=\frac{4-\mu^2z_h^2}{8\pi z_h}.\label{eq:temp}
\end{equation}
We set the horizon at $z_h=1$. In this charged black hole background, we add $e^{-i\omega t}A_x(z)$ and $e^{-i\omega t}g_{tx}(z)$ as
perturbations. The linearized equation of motion for $A_x$ is
\begin{equation}
A_x''+\left(\frac{f'}{f}+\frac{1}{z}\right)A_x'+\left(\frac{\omega^2}{f^2}-\frac{\mu^2}{f}\right)A_x=0.\label{eq:Axq}
\end{equation}
We cannot obtain an analytic solution of this equation in general, so we solve it numerically and then obtain the conductivity. We set $\Lambda=1$ in the following, and remember that $\omega{\rm Im}(\sigma)$ can be shifted by a constant. The real and imaginary parts of the conductivity at different chemical potentials are plotted in Fig.~\ref{fig:sq}. As an approximation, Eq.~(\ref{eq:Axq}) with $f(z)=1-z^2$ can be solved in terms of hypergeometric functions. Then taking into account the boundary conditions at the horizon and the boundary, we obtain the conductivity
\begin{equation}
\sigma(\omega)=\frac{i}{2\omega}\bigg[2\gamma+\psi\bigg(\frac{1-\sqrt{1-\mu^2}-i\omega}{2}\bigg) +\psi\bigg(\frac{1+\sqrt{1-\mu^2}-i\omega}{2}\bigg)\bigg].\label{eq:sqp}
\end{equation}
The real part is plotted in Fig.~\ref{fig:sq} as dashed lines, from which we can see that the above analytic result of the conductivity is qualitatively in agreement with the numerical result.

Figure~\ref{fig:sq} shows that the behavior of the conductivity at large frequencies is the same as the case for zero chemical potential, while there are two distinctive features at small frequencies. One is that the real part of the dc limit Re($\sigma_{\rm dc}$) becomes finite. The other is that the imaginary part has a pole, which corresponds to a delta function at $\omega=0$ in the real part. When we lower the temperature, Re($\sigma_{\rm dc}$) decreases. At zero temperature, i.e., when the black hole is extremal, Re($\sigma_{\rm dc}$) becomes zero. The numerical results of the conductivity at finite chemical potential and finite temperature are also obtained in Ref.~\cite{mai10}. We will analytically explain the above features at finite temperature in the following, and put the zero temperature and low frequency case in Sec. 6.

It turns out that the dc limit of the conductivity can be solved analytically. When $\omega=0$, a special
solution of Eq.~(\ref{eq:Axq}) is $A_x=f'(z)/z$. Thus, the general solution can be expressed as
\begin{equation}
A_x=\frac{f'(z)}{z}\left(C_1+C_2\int_0^z\frac{xdx}{f(x)f'(x)^2}\right).\label{eq:dcAx}
\end{equation}
We impose the incoming-wave boundary condition near the horizon as
\begin{equation}
A_x|_{z\to 1}\sim (1-z)^{2i\omega/(\mu^2-4)}=1+\frac{2i\omega}{\mu^2-4}\ln(1-z)+\cdots.
\end{equation}
By evaluating Eq.~(\ref{eq:dcAx}) near $z=1$ and matching the above boundary condition, we obtain a relation
between $C_1$ and $C_2$ as
\begin{equation}
\frac{C_2}{C_1}=\frac{i}{4}(\mu^2-4)^2\omega+{\cal O}(\omega^2).
\end{equation}
By evaluating Eq.~(\ref{eq:dcAx}) near the boundary $z=0$, we obtain
\begin{equation}
A_x|_{z\to 0}=(\frac{\mu^2}{2}-2+\mu^2\ln z)C_1-\frac{1}{\mu^2}C_2+\cdots.
\end{equation}
Consequently, we obtain the conductivity
\begin{equation}
\sigma=\frac{i}{\omega}\frac{\mu^2-4}{2\mu^2}+\frac{(\mu^2-4)^2}{4\mu^4}+{\cal O}(\omega).\label{eq:dcsig2}
\end{equation}
The $1/\omega^2$ term in Eq.~(\ref{eq:smallo}) vanishes, and there is no divergent term in the real part. The
$1/\omega$ pole in the imaginary part corresponds to a delta function in the real part by the Kramers-Kronig
relation. This infinity of Re($\sigma$) at $\omega=0$ is due to the translational invariance of the system,
and thus it does not correspond to the superconducting phase, as pointed out in Ref.~\cite{har08b}. The FGT sum
rule states that $\int{\rm Re}(\sigma)d\omega$ is independent of the temperature. This can be qualitatively
explained as follows. When we raise the chemical potential, or lower the temperature, the dc limit decreases.
The missing area of $\int{\rm Re}(\sigma)d\omega$ is compensated by the increase of the strength of the delta
function, which corresponds to the residue of the imaginary part. Furthermore, we have numerically checked that
the sum rule is indeed obeyed here.

To recover the explicit temperature dependence, we do not fix the horizon at $z_h=1$, and the result is
\begin{equation}
\sigma=\frac{i}{\omega}\frac{\mu^2z_h^2-4}{2\mu^2z_h^2}+\frac{(\mu^2z_h^2-4)^2}{4\mu^4z_h^3}+{\cal O}(\omega),
\end{equation}
where $z_h$ is solved from Eq.~(\ref{eq:temp}):
\begin{equation}
z_h=\frac{2\sqrt{4\pi^2T^2+\mu^2}-4\pi T}{\mu^2}.
\end{equation}
From dimensional analysis, $[\sigma]=-1$, $[\mu]=1$, and $[T]=1$, we can write the conductivity as
\begin{equation}
{\rm Re}(\sigma_{\rm dc})=\frac{1}{\mu}\Phi(t),\label{eq:dcsig2re}
\end{equation}
where $t\equiv 2\pi T/\mu$ is dimensionless and
\begin{equation}
\Phi(t)=2t^2(\sqrt{t^2+1}+t).
\end{equation}
We have assumed that $\mu>0$ above. If $\mu<0$, we have to replace $\mu$ with $|\mu|$ and define $t=2\pi
T/|\mu|$. The qualitative behavior at low and high temperatures are that $\sigma\sim T^2$ for $T/\mu<<1$, and
$\sigma\sim T^3$ when $T/\mu>>1$.

For comparison, we also give the dc limit of the conductivity for the charged AdS black hole in arbitrary dimension $d\geq 3$. This is the hydrodynamical limit with a chemical potential. The $d=3$ case is studied in Ref.~\cite{har07}, and the $d=4$ case is studied in Ref.~\cite{mye09}. A related work is Ref.~\cite{jai10}. The solution for the charged AdS$_{d+1}$ black hole is in Appendix A. A crucial observation is that $A_x=f'/z^{d-1}$ is a special solution of the perturbation equation (\ref{eq:pertAx}) with $\omega=0$. It was found for AdS$_4$ first \cite{har07}. Then the general solution is
\begin{equation}
A_x=\frac{f'}{z^{d-1}}\left(C_1+C_2\int_0^z\frac{x^{3d-5}dx}{f(x)f'(x)^2}\right),\label{eq:dcAxd}
\end{equation}
where $f(z)=1-(1+\tilde{\mu}^2)z^d+\tilde{\mu}^2z^{2(d-1)}$. This is the $d\geq 3$ counterpart of
Eq.~(\ref{eq:dcAx}). The ratio $C_2/C_1$ can be obtained from the boundary condition near the horizon:
\begin{equation}
\frac{C_2}{C_1}=i\omega[(d-2)\tilde{\mu}^2-d]^2+{\cal O}(\omega^2).
\end{equation}
By performing the integration near $z=0$, we obtain
\begin{equation}
A_x=-d(1+\tilde{\mu}^2)C_1+\left[2(d-1)\tilde{\mu}^2C_1-\frac{C_2}{d(d-2)(1+\tilde{\mu}^2)}\right]z^{d-2}+\cdots.
\end{equation}
By definition Eq.~(\ref{eq:greenab}), we obtain the conductivity
\begin{equation}
\sigma=\frac{i}{\omega}\frac{2(d-1)\tilde{\mu}^2}{d(d-2)(1+\tilde{\mu}^2)}+\left[\frac{(d-2)\tilde{\mu}^2-d}{d(d-2)(1+\tilde{\mu}^2)}\right]^2.
\label{eq:dcsigd}
\end{equation}

To recover the explicit temperature dependence, we need to not fix $z_h=1$, and solve $z_h$ from
Eq.~(\ref{eq:tempd}). From dimensional analysis, we know the conductivity can be written in the form
\begin{equation}
{\rm Re}(\sigma_{\rm
dc})=\frac{1}{z_h^{d-3}}\left[\frac{(d-2)\tilde{\mu}^2z_h^2-d}{d(d-2)(1+\tilde{\mu}^2z_h^2)}\right]^2=\left(\frac{\tilde{\mu}}{d}\right)^{d-3}\Phi(t),
\label{eq:dcsigdre}
\end{equation}
where $t\equiv 2\pi T/\tilde{\mu}$ is dimensionless. The function $\Phi(t)$ is
\begin{equation}
\Phi(t)=\frac{t^2[\sqrt{t^2+d(d-2)}+t]^{d-3}}{[(d-1)\sqrt{t^2+d(d-2)}-t]^2}.
\end{equation}
Again we have assumed that $\tilde{\mu}>0$ above; otherwise we use $|\tilde{\mu}|$. The qualitative behavior at
low and high temperatures are $\sigma\sim T^2$ for $T/\tilde{\mu}<<1$, and $\sigma\sim T^{d-1}$ when
$T/\tilde{\mu}>>1$. Note that $\tilde{\mu}$ is dependent on $d$ as in Eq.~(\ref{eq:mut}). If we make the
replacement as in Eq.~(\ref{eq:rep}) and take the $d\to 2$ limit of Eq.~(\ref{eq:dcsigdre}), we can obtain
Eq.~(\ref{eq:dcsig2re}). This again shows that the conductivity defined by Eq.~(\ref{eq:green}) can be regarded
as the continuum limit of our familiar case $d>2$. The large $t$ behaviors are not the same because the $d\to 2$ limit and the $t\to\infty$
limit do not commute.

If the chemical potential is zero, the dc limit of the conductivity is divergent. In this case, the integration
in Eq.~(\ref{eq:dcAx}) can be done directly and we obtain
\begin{equation}
A_x=-2\Big[C_1+\frac{1}{4}C_2\Big(\ln x-\frac{1}{2}\ln(1-x)-\frac{1}{2}\ln(1+x)\Big)\Big]_0^z.
\end{equation}
By matching the boundary condition at the horizon, we obtain $C_2/C_1=4i\omega+{\cal O}(\omega^2)$. The $\omega\to 0$ limit of the conductivity is
\begin{equation}
\sigma=-\frac{1}{i\omega}\cdot\frac{4C_1}{C_2}=\frac{1}{\omega^2},
\end{equation}
which is the leading order of Eq.~(\ref{eq:smallo}) ($2\pi T=1$).

\section{Superconducting phase}
The superconducting phase corresponds to a charged AdS$_3$ black hole with scalar hair, which exists below a
critical temperature. We work in the probe limit, which means that the Maxwell field and the scalar field do
not backreact on the metric. In the AdS$_{d+1}$ black hole background, the equations of motion for
$\Psi=\psi(z)$ and $A_t=\phi(z)$, and the perturbation equation for $A_x(z)$ are
\begin{eqnarray}
\psi''+\left(\frac{f'}{f}-\frac{d-1}{z}\right)\psi'+\frac{\phi^2}{f^2}\psi-\frac{m^2}{z^2f}\psi &=& 0,\\
\phi''-\frac{d-3}{z}\phi'-\frac{2\psi^2}{z^2f}\phi &=& 0,\\
A_x''+\left(\frac{f'}{f}-\frac{d-3}{z}\right)A_x'+\left(\frac{\omega^2}{f^2}-\frac{2\psi^2}{z^2f}\right)A_x &=&
0,\label{eq:Axsc}
\end{eqnarray}
where $f(z)=1-z^{d}$. The potential for the scalar field is $V(\psi)=m^2\psi^2$, where $\psi$ is real. In the
expansion of $\psi(z)$ near the boundary $z=0$, the coefficients of the leading and next to leading terms are
$\psi^{(1)}$ and $\psi^{(2)}$, respectively. For $-d^2/4\le m^2\le-d^2/4+1$, we can choose any of $\psi^{(i)}$
($i=1,2$) to condense. For $m^2\ge -d^2/4+1$, we can only take $\psi^{(2)}$ to condense.

Now we study the $d=2$ case. We take the mass of the scalar equal to the BF bound $m_{\rm BF}^2=-1$. At the horizon $z=1$, the boundary conditions are $\phi|_{z=1}=0$ \cite{har08a}, $\psi=2\psi'$, and $A_x|_{z\to 1}\sim(1-z)^{-i\omega/2}+\cdots$. The asymptotic behavior of the various fields near the boundary $z=0$ is
\begin{eqnarray}
\psi &=& \psi^{(1)}z\ln z+\psi^{(2)}z+\cdots,\label{eq:psiz}\\
\phi &=& \mu\ln z+\rho+\cdots,\\
A_x &=& {\cal A}\ln z+{\cal B}+\cdots.\label{eq:sig0}
\end{eqnarray}
The Green's function for $A_x$ is obtained by
\begin{equation}
G=-\frac{\cal B}{\cal A}=-\lim_{z\to 0}\frac{A_x-A_x'z\ln z}{A_x'z}.
\end{equation}

\begin{figure}
\centering
\begin{minipage}[c]{0.5\textwidth}
\centering
\includegraphics[width=0.90\textwidth]{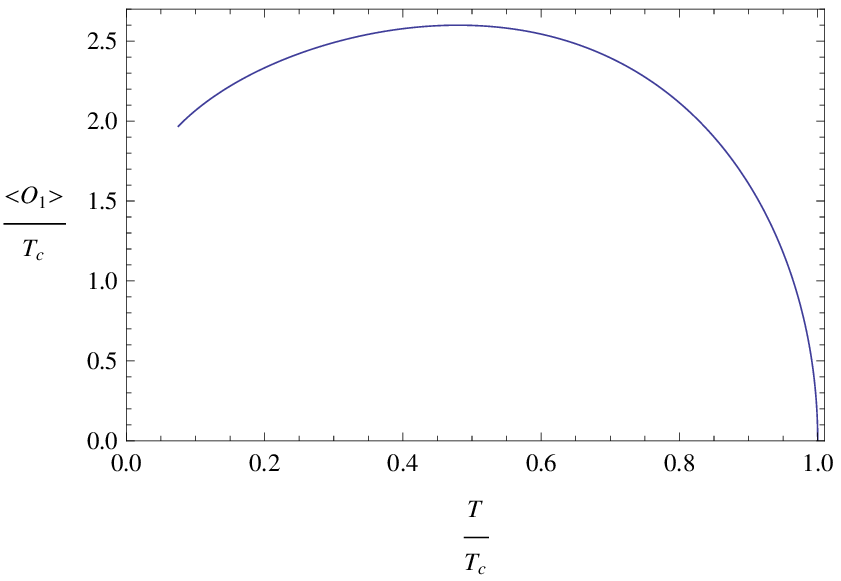}
\end{minipage}%
\begin{minipage}[c]{0.5\textwidth}
\centering
\includegraphics[width=0.95\textwidth]{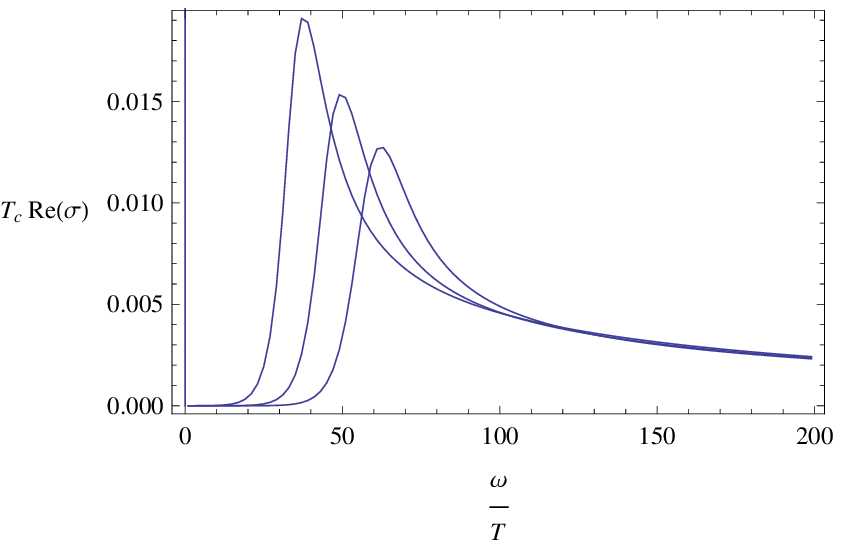}
\end{minipage}
\caption{\label{fig:O1sig} The left panel is the condensate of the operator $\langle{\cal O}_1\rangle$ as a
function of temperature, when $m^2=m_{\rm BF}^2$. The right panel is the real part of the conductivity as a
function of frequency at low temperatures $T/T_c=0.27$, 0.20, 0.16 (from left to right).}
\end{figure}

\begin{figure}
\centering
\begin{minipage}[c]{0.5\textwidth}
\centering
\includegraphics[width=0.90\textwidth]{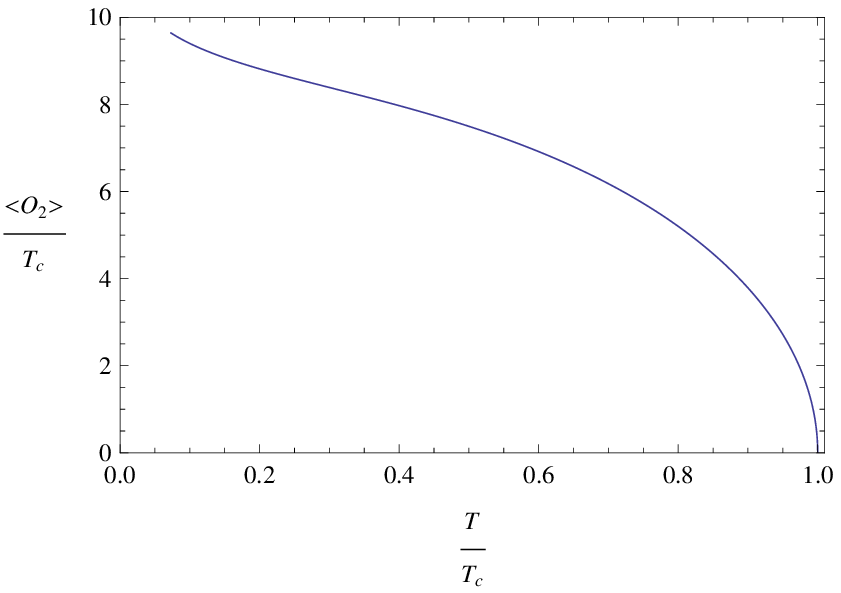}
\end{minipage}%
\begin{minipage}[c]{0.5\textwidth}
\centering
\includegraphics[width=0.95\textwidth]{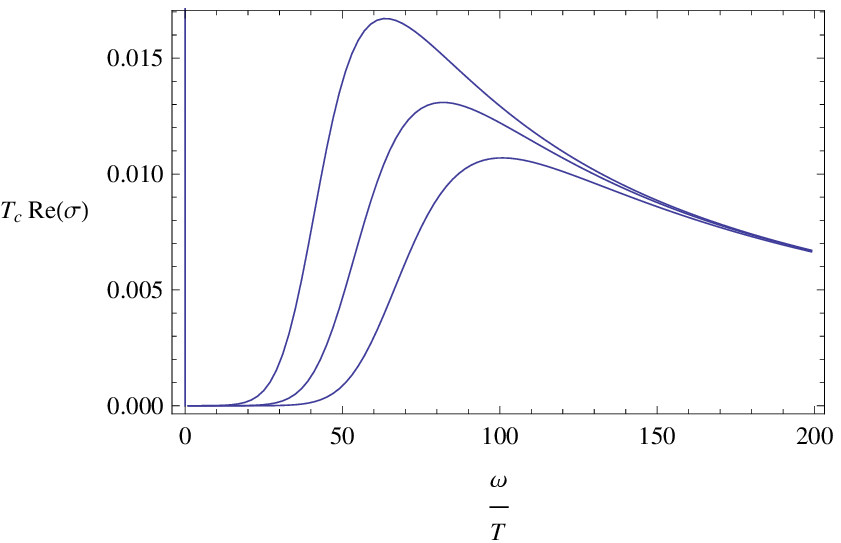}
\end{minipage}
\caption{\label{fig:O2sig} The left panel is the condensate of the operator $\langle{\cal O}_2\rangle$ as a
function of temperature, when $m^2=m_{\rm BF}^2$. The right panel is the real part of the conductivity as a
function of frequency at low temperatures $T/T_c=0.21$, 0.17, 0.14 (from left to right).}
\end{figure}

To achieve the superconducting phase, we need a spontaneous symmetry breaking. We expect that an operator dual
to the scalar field condenses without being sourced. There are two ways to choose the scalar operator and the
boundary condition:
\begin{equation}
\langle{\cal O}_i\rangle=\psi^{(i)},\qquad\epsilon_{ij}\psi^{(j)}=0,\qquad i=1,2.
\end{equation}
Note that the $\psi^{(2)}=0$ theory is not precisely conformal \cite{wit01,mar06}. The condensates of $\langle{\cal O}_1\rangle$ and $\langle{\cal O}_2\rangle$ together with the conductivities are plotted in Figs.~\ref{fig:O1sig} and \ref{fig:O2sig}, respectively. By fitting these curves as
$\langle{\cal O}\rangle=a(1-T/T_c)^b$, where $T_c$ is the critical temperature, we obtain $b\approx 0.5$ for both cases as $T\to T_c$. This implies that a second order phase transition occurs. We find
\begin{equation}
\langle{\cal O}_1\rangle\approx 5.3T_c(1-T/T_c)^{1/2},
\end{equation}
where $T_c\approx 0.050\mu$, and
\begin{equation}
\langle{\cal O}_2\rangle\approx 12.2T_c(1-T/T_c)^{1/2},
\end{equation}
where $T_c\approx 0.136\mu$. Below a finite frequency $\omega_g$, the real part of the conductivity falls off exponentially. By fitting the curves, we obtain ${\rm Re}(\sigma)\sim e^{(\omega-\omega_g)/T}$, where $\omega_g/T_c\approx 9$.


\begin{figure}
\centering
\begin{minipage}[c]{0.5\textwidth}
\centering
\includegraphics[width=0.95\textwidth]{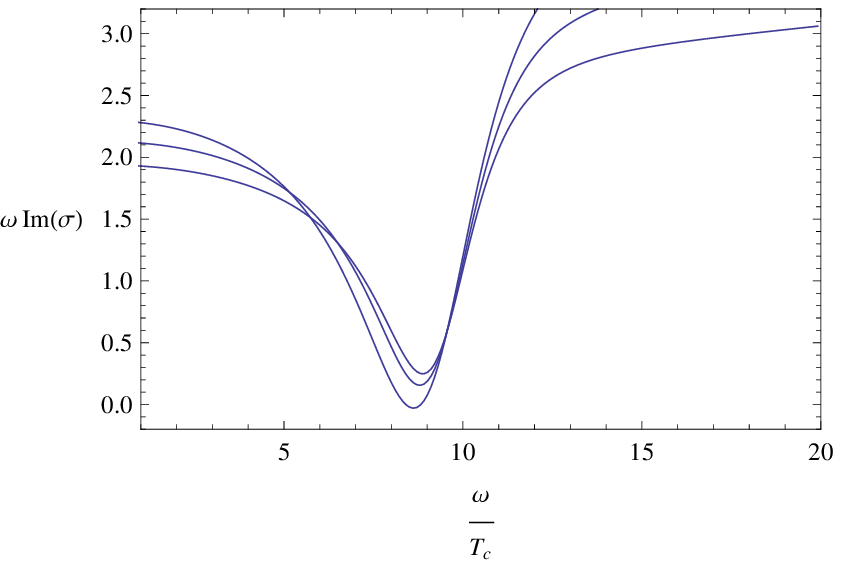}
\end{minipage}%
\begin{minipage}[c]{0.5\textwidth}
\centering
\includegraphics[width=0.95\textwidth]{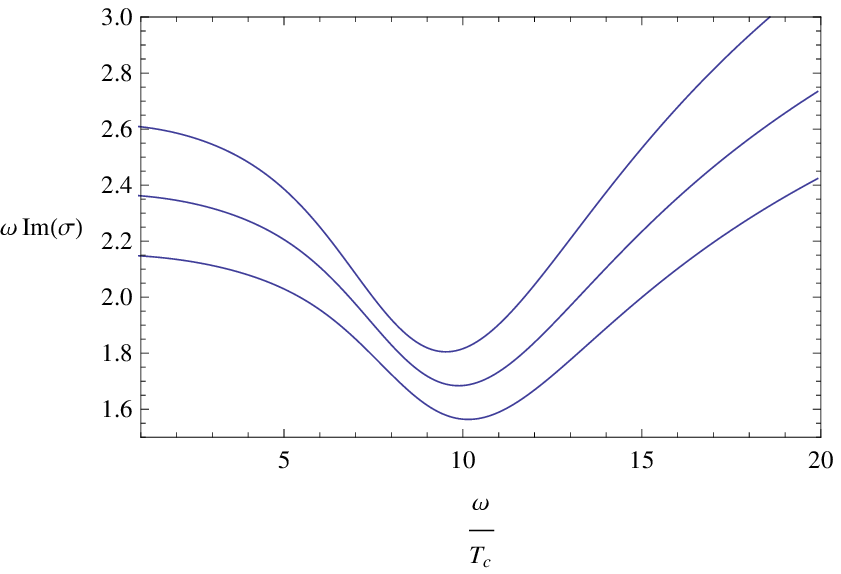}
\end{minipage}
\caption{\label{fig:pole} The behavior of $\omega{\rm Im}(\sigma)$ as a function of frequency when
$\langle{\cal O}_1\rangle$ or $\langle{\cal O}_2\rangle$ condensates. It approaches to a constant as $\omega\to
0$, which implies that there are a pole in Im($\sigma$) and a delta function in Re($\sigma$) at $\omega=0$. The
value $\omega_g/T_c$ can also be found in these plots.}
\end{figure}

The behavior of $\omega{\rm Im}(\sigma)$ is plotted in Fig.~\ref{fig:pole}, from which we can see that there is a pole in the imaginary part of the conductivity. The Kramers-Kronig relation
\begin{equation}
{\rm Im}[\sigma(\omega)]=-\frac{1}{\pi}{\cal P}\int_{-\infty}^\infty\frac{{\rm
Re}[\sigma(\omega')]}{\omega'-\omega}d\omega'
\end{equation}
implies
\begin{equation}
{\rm Im}[\sigma(\omega)]=\frac{n_s}{\omega}\quad\Leftrightarrow\quad{\rm Re}[\sigma(\omega)]=\pi
n_s\delta(\omega).
\end{equation}
Therefore, there is a delta function at $\omega=0$ in the real part. Together with the frequency gap $\omega_g$ in Figs.~\ref{fig:O1sig} and \ref{fig:O2sig}, this is the feature of a superconducting phase transition when $T<T_c$. We can see that $\omega_g/T_c\approx 9$. The FGT sum rule states that $\int{\rm Re}(\sigma)d\omega$ is independent of the temperature, which implies that
\begin{equation}
\left[2\int_{0^+}^\infty{\rm Re}[\sigma(\omega)]d\omega+\lim_{\omega\to 0}\omega{\rm
Im}[\sigma(\omega)]\right]_{T_1}^{T_2}=0.
\end{equation}
The missing area when we lower the temperature is exactly compensated by the residue of the pole in Im[$\sigma(\omega)$]. The sum rule can be regarded as a check that our program for numerical calculation is correct.

\begin{figure}
\centering
\begin{minipage}[c]{0.5\textwidth}
\centering
\includegraphics[width=0.90\textwidth]{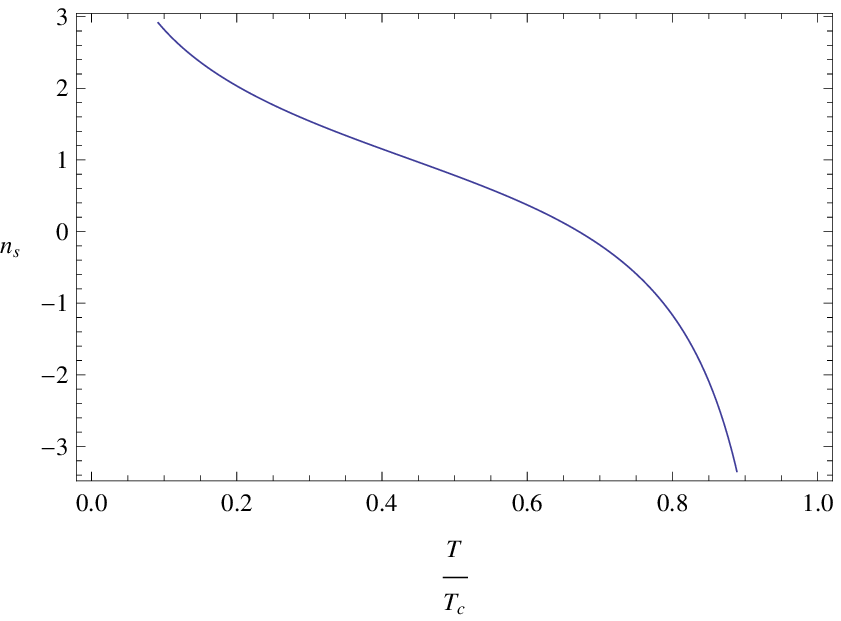}
\end{minipage}%
\begin{minipage}[c]{0.5\textwidth}
\centering
\includegraphics[width=0.95\textwidth]{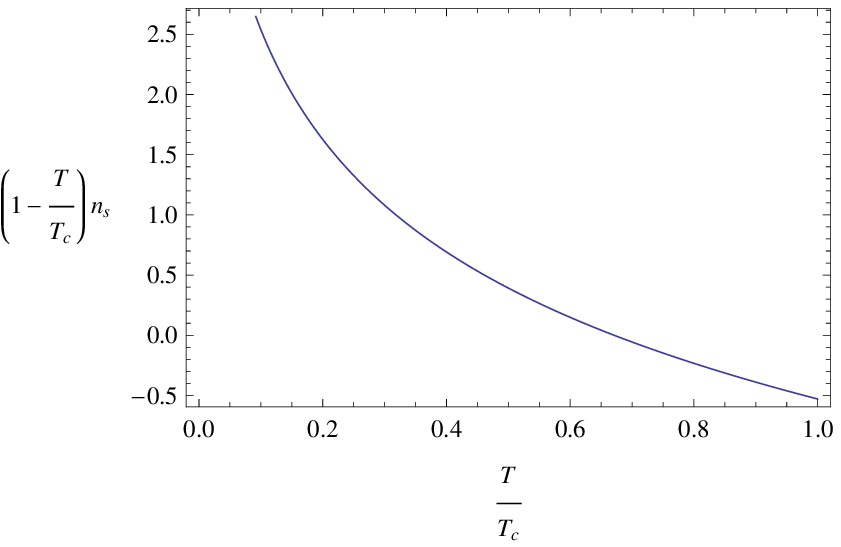}
\end{minipage}
\caption{\label{fig:ns} Behavior of the superfluid density $n_s$ as a function of temperature, when the
operator $\langle{\cal O}_2\rangle$ condensates and $m^2=m_{\rm BF}^2$.}
\end{figure}

A distinctive feature is that the superfluid density is divergent when $T\to T_c$, which is shown in
Fig.~\ref{fig:ns}. For the condensate of $\langle{\cal O}_2\rangle$, we find that the behavior of the superfluid density is about
\begin{equation}
n_s\sim -0.53(1-T/T_c)^{-1}+1.3.
\end{equation}
A rough explanation is as follows. When $T\to T_c$, the superconducting phase approaches the normal phase. Near the boundary, $\psi$ behaves as $\psi=\langle{\cal O}_2\rangle z$. By comparing the perturbation equations for the two phases, we can see that $2\psi^2/z^2=2\langle{\cal O}_2\rangle^2$ in Eq.~(\ref{eq:Axsc}) corresponds to $\mu^2$ in Eq.~(\ref{eq:Axq}). From Eq.~(\ref{eq:dcsig2}), we can see that ${\rm Res}[{\rm Im}(\sigma)]\sim 1/\mu^2$ as $\mu\to 0$. Therefore, $n_s\sim 1/\langle{\cal O}_2\rangle^2\sim(1-T/T_c)^{-1}$ as $T\to T_c$.

If $d\geq 3$, we obtain ${\rm Res}[{\rm Im}(\sigma)]\sim\mu^2$ as $\mu\to 0$ from Eq.~(\ref{eq:dcsigd}). Therefore, $n_s\sim 1-T/T_c$ as $T\to T_c$. Furthermore, the equations are well-defined even if $d$ is not an integer. We try $d=2.05$ and obtain $n_s\to 0$ as $T\to T_c$. This implies that the divergence can be regularized by the dimension $d=2+\epsilon$.

\begin{figure}
\centering
\begin{minipage}[c]{0.5\textwidth}
\centering
\includegraphics[width=0.90\textwidth]{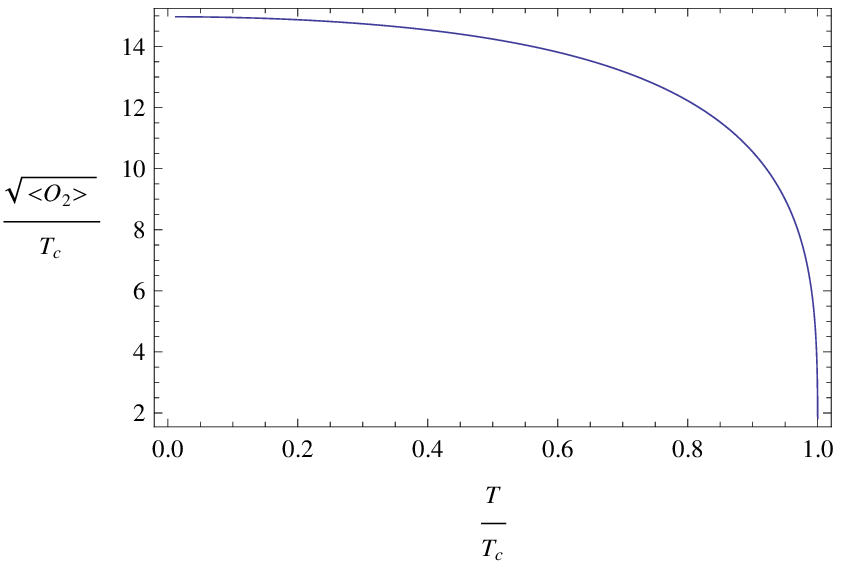}
\end{minipage}%
\begin{minipage}[c]{0.5\textwidth}
\centering
\includegraphics[width=0.95\textwidth]{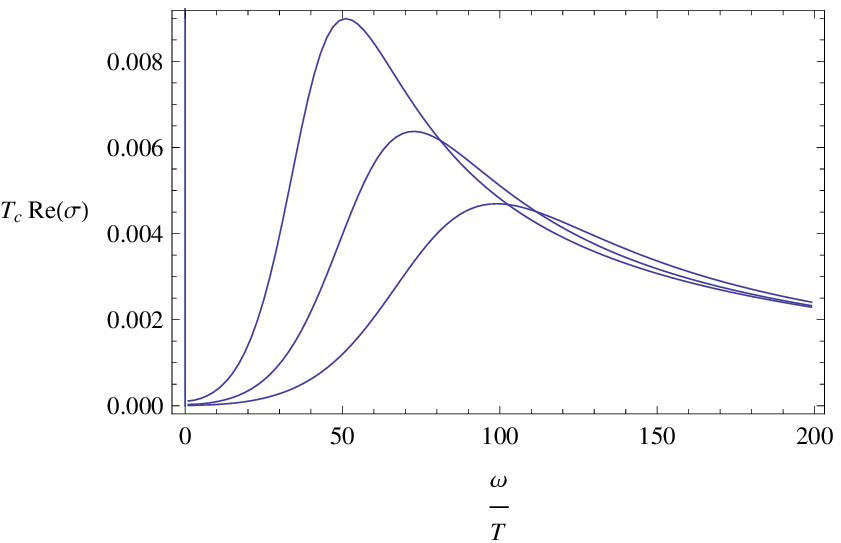}
\end{minipage}
\caption{\label{fig:Osig} The left panel is the condensate of the operator $\langle{\cal O}_2\rangle$ as a function of
temperature, when $m^2=0$. The right panel is the real part of the conductivity as a function of frequency at
low temperatures $T/T_c=0.23$, 0.16, 0.12 (from left to right).}
\end{figure}

We take the mass of the scalar field be zero, and can also obtain the superconducting phase. When $m=0$, the
asymptotic behavior of the scalar field near the boundary becomes
\begin{equation}
\psi=\psi^{(1)}+\psi^{(2)}z^2+\cdots.
\end{equation}
In this case we can only choose the scalar operator by $\langle{\cal O}_2\rangle=\psi^{(2)}$, with the boundary
condition $\psi^{(1)}=0$. The condensate of $\langle{\cal O}_2\rangle$ together with the conductivity is shown
in Fig.~\ref{fig:Osig}. By fitting the curve near $T_c$, we obtain
\begin{equation}
\langle{\cal O}_2\rangle\approx 365T_c^2(1-T/T_c)^{1/2},
\end{equation}
where $T_c\approx 0.046\mu$.

\section{Away from the probe limit}
For AdS$_4$/CFT$_3$, the holographic superconductors away from the probe limit have been studied in
Ref~\cite{har08b}, and the zero temperature limit has been studied in Refs.~\cite{gub09,hor09}. The equations of
motion from the action Eq.~(\ref{eq:action}) can be found in Ref.~\cite{har08b}. We want to know the properties
for AdS$_3$, i.e., $d=2$, and we start with arbitrary $d$ for generality. To consider the backreaction of the
Maxwell field and the scalar field on the spacetime, we take the metric ansatz
\begin{equation}
ds^2=-h(r)e^{-\chi(r)}dt^2+\frac{dr^2}{h(r)}+r^2\sum_{i=1}^{d-1}dx_i^2,
\end{equation}
together with
\begin{equation}
A=\phi(r)dt,\qquad \psi=\psi(r).
\end{equation}
The equation of motion for the scalar field is
\begin{equation}
\psi''+\left(\frac{h'}{h}-\frac{\chi'}{2}+\frac{d-1}{r}\right)\psi'+\frac{q^2\phi^2e^\chi}{h^2}\psi-\frac{1}{2h}V'(\psi)=0.\label{eq:psi}
\end{equation}
Maxwell equations give
\begin{equation}
\phi''+\left(\frac{\chi'}{2}+\frac{d-1}{r}\right)\phi'-\frac{2q^2\psi^2}{h}\phi=0.
\end{equation}
Einstein equations give
\begin{eqnarray}
(d-1)\chi'+2r\psi'^2+\frac{2rq^2\phi^2\psi^2e^\chi}{h^2} &=& 0,\\
(d-1)h'+(d-1)\left(\frac{d-2}{r}-\frac{\chi'}{2}\right)h+\frac{r\phi'^2e^\chi}{2}-\frac{d(d-1)r}{L^2}+rV(\psi)
&=& 0.
\end{eqnarray}
The above equations of motion are valid even if $d=1$, i.e., the AdS$_2$ case.

Then we add $e^{-i\omega t}A_x(z)$ and $e^{-i\omega t}g_{tx}(z)$ as perturbations. The perturbation equation
for $A_x$ is
\begin{equation}
A''_x+\left(\frac{h'}{h}-\frac{\chi'}{2}+\frac{d-3}{r}\right)A'_x+\left(\frac{\omega^2}{h^2}e^\chi-\frac{2q^2\psi^2}{h}\right)A_x
=\frac{\phi'}{h}e^\chi\left(-g'_{tx}+\frac{2}{r}g_{tx}\right).\label{eq:fullAx}
\end{equation}
The perturbation equation for $g_{tx}$ is independent of the dimension:
\begin{equation}
g'_{tx}-\frac{2}{r}g_{tx}+\phi'A_x=0.
\end{equation}
By substituting the above equation into Eq.~(\ref{eq:fullAx}), we obtain
\begin{equation}
A''_x+\left(\frac{h'}{h}-\frac{\chi'}{2}+\frac{d-3}{r}\right)A'_x+\left[\left(\frac{\omega^2}{h^2}
-\frac{\phi'^2}{h}\right)e^\chi-\frac{2q^2\psi^2}{h}\right]A_x=0.\label{eq:Axfull}
\end{equation}
After a change of variables
\begin{equation}
dz=\frac{e^{\chi/2}}{h}dr,\qquad\tilde{A}_x=r^{(d-3)/2}A_x,\label{eq:chvar}
\end{equation}
Eq.~(\ref{eq:Axfull}) becomes a Schr\"{o}dinger-like equation:
\begin{equation}
-\tilde{A}_{x,zz}+\tilde{V}(z)\tilde{A}_x=\omega^2\tilde{A}_x.\label{eq:schr}
\end{equation}
The method on how to make the change of variables is reviewed in Appendix C of Ref.~\cite{her07b}. The potential is
\begin{equation}
\tilde{V}(z)=h(\phi_{,r}^2+2q^2\psi^2e^{-\chi})+\frac{(d-3)h^2}{2e^\chi}\left[\frac{d-5}{2r^2}
+\frac{1}{r}\left(\frac{h_{,r}}{h}-\frac{\chi_{,r}}{2}\right)\right].\label{eq:V}
\end{equation}
It has been proved that the first part vanishes both at the boundary for certain scaling dimensions of $\psi$ and at the horizon \cite{hor09}. However, the second part is divergent at the boundary when $d\neq 3$, though it vanishes at the horizon. In AdS$_4$, the conductivity can be determined by the reflection coefficient of the potential barrier with purely ingoing wave at the horizon, which is used to prove that the real part of the conductivity is not strictly zero at low frequency even at zero temperature \cite{hor09}. This method cannot be generalized to other dimensions because of the divergence of the potential when $d\neq 3$. Therefore, it is still an open question whether the real part of the conductivity is strictly zero at low frequency (has a hard gap) when $d\neq 3$.

It was found first in the AdS$_4$ case that the low frequency limit of the conductivity at zero temperature can be obtained by considering the conserved flux ${\cal F}=i(A_x^*\partial_zA_x-{\rm c.c.})$ \cite{gub09,hor09}. If $\tilde{V}\approx V_0/z^2$ near the horizon, then the conductivity is given by ${\rm Re}(\sigma)\sim\omega^\delta$, where $\delta=\sqrt{4V_0+1}-1$ \cite{hor09}. We will show that this conclusion can be generalized to arbitrary dimension $d\geq 2$ straightforwardly. From the Schr\"{o}dinger-like equation (\ref{eq:schr}), we can see that there is a conserved flux
\begin{equation}
{\cal F}=i(\tilde{A}_x^*\partial_z\tilde{A}_x-{\rm c.c.}),
\end{equation}
where c.c. denotes the complex conjugate. From Eq.~(\ref{eq:chvar}), we can see that $r\in[1,\infty)$ and $z\in(-\infty,0]$. Near the boundary, $z=-1/r\to 0$. When $d=2$, we have $\tilde{A}_x=\sqrt{-z}({\cal A}\ln(-z)+{\cal B})$. After discarding the divergent term, we obtain
\begin{equation}
{\cal F}=-i({\cal B}^*{\cal A}-{\rm c.c.})=2\omega|{\cal A}|^2{\rm Re}(\sigma).\label{eq:flux}
\end{equation}
When $d\geq 3$, $\tilde{A}_x=(-z)^{(3-d)/2}({\cal A}+{\cal B}(-z)^{d-2})$, and we still obtain ${\cal F}\sim\omega|{\cal A}|^2{\rm Re}(\sigma)$. If the potential near the horizon $z\to-\infty$ is $\tilde{V}\approx V_0/z^2$, we have
\begin{equation}
\tilde{A}_x\sim\sqrt{-\omega z}H_\nu^{(1)}(-\omega z)\sim\left\{\begin{array}{ll}(-\omega z)^{1/2-\nu}, & \qquad |z|\gg 1,\;\omega|z|\ll 1~({\rm outer~region})\\
e^{-i\omega z}, & \qquad |z|\gg 1,\;\omega|z|\gg 1~({\rm inner~region})\end{array}\right.
\end{equation}
where $\nu=\sqrt{V_0+1/4}$. By matching the outer region and the boundary, we have ${\cal A}\sim \omega^{1/2-\nu}$. The inner region gives ${\cal F}\sim\omega$. By comparing this with Eq.~(\ref{eq:flux}), we obtain ${\rm Re}(\sigma)\sim\omega^{2\nu-1}$.

The condition $\tilde{V}\approx V_0/z^2$ near the horizon is satisfied for many cases of the holographic superconductors at zero temperature \cite{hor09}, including the W-shaped potential model studied in Ref.~\cite{gub09}. It is also satisfied for the extremal charged black hole, which has an AdS$_2$ in the near horizon region \cite{fau09}. We will take the extremal charged AdS$_3$ black hole as an example. If there is no scalar field, the solution is
\begin{equation}
\chi=\psi=0,\qquad h=r^2-1-\frac{\mu^2}{2}\ln r,\qquad \phi=\mu\ln r.
\end{equation}
The temperature is $T=(4-\mu^2)/8\pi$, so the extremal limit is $\mu=2$. The near horizon limit is
AdS$_2\times\mathbb{R}$ as follows:
\begin{equation}
ds^2=-2(r-1)^2dt^2+\frac{dr^2}{2(r-1)^2}+dx^2.
\end{equation}
The relation between $z$ and $r$ is $z=-1/[2(r-1)]$. The near horizon limit of Eq.~(\ref{eq:V}) gives
$\tilde{V}\approx 8(r-1)^2=2/z^2$. Therefore, ${\rm Re}(\sigma)\sim\omega^2$, which was obtained in Ref.~\cite{mai10}.

At low temperature, instability will cause the black hole to form scalar hair. By defining $\tilde{r}=r-1$, the near horizon limit of Eq.~(\ref{eq:psi}) is a wave equation for AdS$_2$ with a new effective mass:
\begin{equation}
\psi_{,\tilde{r}\tilde{r}}+\frac{2}{\tilde{r}}\psi_{,\tilde{r}}-\frac{m_{\rm eff}^2}{\tilde{r}^2}\psi=0,\qquad
m_{\rm eff}^2=\frac{m^2-2q^2}{2}.
\end{equation}
The BF bound for AdS$_2$ is $-1/4$. Therefore, a sufficient condition for the instability is that $m^2-2q^2<-1/2$, where $m^2>-1$, the BF
bound for AdS$_3$.

\section{Summary}
We have studied the main features of the one-dimensional holographic superconductor by AdS$_3$/CFT$_2$ correspondence. The frequency and temperature dependence of the conductivity can be qualitatively summarized as follows. The dimension of the conductivity for one-dimensional materials is $[\sigma]=-1$. At zero temperature and zero chemical potential, the frequency $\omega$ is the only scale. Therefore, we have $\sigma\sim 1/\omega$, and there can also be a $\ln(\omega/\Lambda)$ term. Generally, the behavior for large $\omega$ is always $\sigma\sim 1/\omega$, if $\omega$ is the only important scale. When we add temperature, the dc limit of the conductivity is still divergent and behaves as $\sigma\sim 2\pi T/\omega^2$. When we further add chemical potential, the real part of the dc limit Re($\sigma_{\rm dc}$) becomes finite and decreases as the temperature is lowered. And Re($\sigma_{\rm dc}$) becomes zero when $T=0$. When we add scalar hair to the system, Re($\sigma$) falls off exponentially at a finite frequency, if the temperature is below a critical temperature. There is a delta function at $\omega=0$ in both normal and superconducting phases, but only the latter one has a spontaneous breaking of the U(1) symmetry.

Analytic solutions of the conductivity calculated from the charged AdS$_{d+1}$ black holes and their special cases can be summarized as follows:
\begin{center}
\begin{tabular}{|l|l|p{1.9cm}|p{1.9cm}|p{1.9cm}|p{1.3cm}|}
\hline
\multicolumn{2}{|c|}{$\sigma(\omega)$} & $d=2$ & $d=3$ & $d=4$ & $d\geq 5$\\
\hline \multicolumn{2}{|l|}{$T=0,\mu=0$} & \multicolumn{4}{|c|}{Eq.~(\ref{eq:sigads})}\\
\hline \multicolumn{2}{|l|}{$T\neq 0,\mu=0$} & Eq.~(\ref{eq:sigT}) & Refs.~\cite{her07a,hor08} & Refs.~\cite{hor08,mye07} & N\\
\hline $T\neq 0,\mu\neq 0$ & $\omega\to 0$ & Eq.~(\ref{eq:dcsig2}) & \multicolumn{3}{|c|}{Eq.~(\ref{eq:dcsigd})}\\
\hline $T=0,\mu\neq 0$ & $\omega\to 0$ & \multicolumn{4}{|c|}{${\rm Re}(\sigma)\sim\omega^\delta$}\\
\hline
\end{tabular}
\end{center}
Some special cases of Eq.~(\ref{eq:dcsigd}) are in Refs.~\cite{har07,mye09}. The references for the last row are in Sec. 6. And N denotes that there is no analytic solution as far as we know. Near the phase transition, if the zero mode of the scalar field can be analytically solved, the scaling properties may be analytically obtained \cite{her10} (the $p$-wave model was considered earlier in Ref.~\cite{her09}). Unfortunately this is not the case for AdS$_3$ due to the $\ln z$ term.

There are some remaining questions, such as why the conductivity does not have spikes as in higher-dimensional cases when $m=m_{\rm BF}$ \cite{hor08}, whether the conductivity has a hard gap at $T=0$, and what is the relation between this model and other (1+1)-dimensional systems in condensed matter physics. The embedding of this model into the superstring or M-theory has not been examined. Other gravity systems may also be used to construct the holographic dual of the one-dimensional superconductor, such as the D1-brane. Many other features in AdS$_4$/CFT$_3$ and AdS$_5$/CFT$_4$ can be studied in AdS$_3$/CFT$_2$ in parallel. For example, one could consider the $p$-wave and $d$-wave superconductors, and the fermionic correlation functions.

\section*{Acknowledgements}
I would like to thank my advisor, Prof. C.P. Herzog, for his guidance. This work was supported in part by the National Science Foundation under Grants No. PHY-0844827 and PHY-0756966.

\appendix
\section{AdS$_3$ as a limit of AdS$_{d+1}$}
The charged AdS$_3$ black hole can be obtained by taking the $d\to 2$ limit from AdS$_{d+1}$ black holes, if we
regard $d$ as a continuous quantity. The solution of the AdS$_{d+1}$ charged black hole system with the ansatz
Eq.~(\ref{eq:ansatz}) and $A=\phi(z)dt$ is
\begin{eqnarray}
f(z) &=& 1-(1+\tilde{\mu}^2)\Big(\frac{z}{z_h}\Big)^d+\tilde{\mu}^2\Big(\frac{z}{z_h}\Big)^{2(d-1)},\label{eq:solf}\\
A(z) &=& \mu\Big[1-\Big(\frac{z}{z_h}\Big)^{d-2}\Big]dt,\label{eq:solA}
\end{eqnarray}
where
\begin{equation}
\tilde{\mu}^2=\frac{d-2}{2(d-1)}\mu^2.\label{eq:mut}
\end{equation}
The Hawking temperature is
\begin{equation}
T=\frac{d-(d-2)\tilde{\mu}^2z_h^2}{4\pi z_h}.\label{eq:tempd}
\end{equation}
Firstly we replace $\mu$ with $\mu'$ as follows
\begin{equation}
\mu\to\mu'=-\frac{\mu}{d-2},\qquad \tilde{\mu}^2\to\tilde{\mu}'^2=\frac{\mu^2}{2(d-1)(d-2)}.\label{eq:rep}
\end{equation}
Then after we take the $d\to 2$ limit, use $z^\epsilon=1+\epsilon\ln z+\cdots$, and discard the $1/\epsilon$
divergence, Eqs.~(\ref{eq:solf}) and (\ref{eq:solA}) are exactly the AdS$_3$ solution.

\end{document}